\documentstyle[prl,aps,epsf,twocolumn]{revtex}
\begin{document}
\pagestyle{empty}
\draft
\twocolumn
[
\hsize\textwidth\columnwidth\hsize\csname @twocolumnfalse\endcsname
]

{\bf Comment on "Phenomenological model of 

\noindent
Dynamic Nonlinear Response in Relaxor

\noindent
Ferroelectrics"}

~

In the recent letter[1]  Glazounov and Tagantsev (GT) proposed the phenomenological model of dynamic response of relaxors. The model is based on the well known  Landau-Khalatnikov (LK) equation for the  polarization dynamics $\partial  P /\partial t = -(1/\eta) \partial F /\partial P, ~ F(P) =(1/2)\alpha P^2 +(1/4) \beta P^4 -PE$. The new element introduced by  the authors is the averaging of the solution of LK equation over the broad distribution of relaxation times $\tau= \eta/\alpha $ which, as the authors assume, resembles the  low frequency dispersion of the dielectric response observed in relaxors. The purpose of the present comment is to clarify the physical assumptions involved in the  GT approach and show that  models  based on LK equation can not describe properly the   dynamic response  in  relaxors ferroelectrics.

Note first  that LK equation completely ignores all the inhomogeneities of the system both temporal and spatial, and the relaxation time $\tau$ is  characteristic of the dynamics of the long wave length component of polarization  with zero wave vector. The legitimate question to ask in this case is  what is the meaning of the averaging over $\tau$  proposed by GT? The following scenario might be true. Let us  assume that  that the crystal is divided  into macroregions within which the coefficient $\tau$ has the same magnitude. Then  the average over $\tau$  just means the average of the polarization response for different macroregions. However, the above physical picture is very far from  the  real situation taking place in relaxors, where the existence of nanoscale dynamical  inhomogeneities(polar clusters)  are well documented.  The concept of nanoscale polar clusters provides  more natural explanation of  the broad distribution of relaxation times in relaxors than that implied by the GT approach.  Indeed, each polar cluster may be characterized but its own value of $\tau$ with  the correlations between the values of $\tau$ for different clusters being negligibly small.  A formalism which explicitly takes into  account the dynamics of interacting polar clusters in relaxors  has been developed in Ref[2] and applied successfully to the dielectric experiments in PST and  PMN.

We will illustrate below that  GT approach leads to the values of the dielectric response which in general is inconsistent with the experiments in relaxors. As follows from GT model, the linear dielectric permittivity may be written in the form 
\begin{equation}
\epsilon'(\omega, T) = \epsilon_s(T) (1-q(\omega, T))
\label {GT}
\end{equation}
where $1 - q(\omega,T)= \langle 1/(1+\omega^2 \tau^2)\rangle$, and  $\langle...\rangle$ denotes the average over $\tau$. It is apparent that the quantity $q(\omega, T)$ is the fraction of polar regions which are effectively frozen on the time scale $\omega^{-1}$.   Indeed, if for the major part of polar regions $\omega\tau <<1$, then  $q \rightarrow 0$, and  in the opposite limit   $\omega\tau >>1$   $q \rightarrow 1$. Since relaxor ferroelectrics are close to ferroelectric instability they are characterized by a strong  increase of the static (field cooled) permittivity $\epsilon_s(T)$ with the temperature decrease. At the same time the frequency dependent permittivity $\epsilon^{\prime}(\omega,T)$ decreases with  temperature below a well defined temperature $T_m$.  The applicability of Eq.(1) in this situation assumes that  the increase of $\epsilon_s$ is  compensated for by an increase of the parameter $q(\omega,T)$. However, recent NMR experiments [3] which obtain direct information on the magnitude of $q(\omega,T)$ from measurements of the inhomogeneous NMR line width do not support this assumption. Indeed, it was found[3] that in PMN, $q \approx 0.2$ at $\omega=10^5 Hz$ and T=220K which, accordingly to Eq.(1) leads to the conclusion that $\epsilon^{\prime}(\omega)/\epsilon_s \approx 0.8$. However, dielectric experiments[4] performed in the same sample give a much lower ratio $\epsilon^{\prime}(\omega)/\epsilon_s  <0.05$ for the same values of the parameters.

This example, accompanied by the given above physical picture of the meaning of the averaging of the solution of the LK equation,  clearly shows that GT phenomenological model is inconsistent with the mechanism of relaxor behavior since it does not take into account the existence of short range polar clusters.
Therefore, one can doubt  the validity for relaxors of the obtained in Ref.[1] expression    
 $P_3^{\prime} \sim \beta \epsilon^{\prime 3} (\omega)\epsilon'(3 \omega)$  for nonlinear third order polarization response $P_3^{\prime}$ with $\epsilon^{\prime}(\omega)$ given by Eq.(1).  Note, however, that GT employed the above expession for $P_3^{\prime}$
only for evaluating  the coefficient $\beta$  using experimental values for  both $\epsilon^{\prime}(\omega)$ and $P_3^{\prime}$. In order to test how sensitive is such a procedure to the particular model we performed the same calculations using the  interacting polar clusters model[2]. We obtained

\begin{equation}
P_3^{\prime} \sim \beta \epsilon^{\prime 3} (\omega)\epsilon'(3 \omega) (1-q(3\omega))/(1-q(\omega)).
\end{equation}
Since in relaxors the frequency dependence $q$ is rather weak,  such that  $q( 3 \omega) \approx q( \omega)$, one can state that the evaluation of the coefficient $\beta$ given by GT is quite reliable, although almost with the same degree of accuracy one could have used the expression $P_3^{\prime} \sim \beta \epsilon^{\prime 4}(\omega)$ employed earlier in Ref.[5].

~

\noindent
1. A. E. Glazounov and A.K. Tagantsev, PRL {\bf 85}, 2192

(2000).

\noindent
2. B. E. Vugmeister and H.Rabitz, Phys.Rev. B {\bf 57}, 7581

(2000); {\bf 61}, 14448 (2000).

\noindent
3.R.Blinc, {\it et al}., Phys.Rev.Let. {\bf 83}, 424 (1999).

\noindent
4. A. Levstic, {\it et al}., Phys.Rev. B {\bf 57}, 11204 (1998).   

\noindent
5.V. Bobnar, {\it et al}., Phys.Rev. Lett. {\bf 84}, 5892 (2000). 

\begin{center}
B.E. Vugmeister and H. Rabitz

{\it Department of Chemistry Princeton University

Princeton, NY 08544 }
\end{center}
\end{document}